\font\tx=cmr10 at 11pt  \font\ma=cmmi10 at 11pt  \font\sy=cmsy10 at 11pt
\textfont0=\tx          \textfont1=\ma           \textfont2=\sy
\font\sub=cmr8          \font\masub=cmmi8        \font\sysub=cmsy8
\scriptfont0=\sub       \scriptfont1=\masub      \scriptfont2=\sysub
\font\title=cmbx12      \font\author=cmr12       \font\bx=cmbx10 at 11pt
\font\it=cmti10 at 11pt
\baselineskip=13pt      \parindent=18pt          \raggedbottom
\lefthyphenmin=3        \righthyphenmin=4        \hyphenpenalty=200
\def\space{\vskip 13pt}
\def\section#1{\space\space\goodbreak\leftline{\bx#1}\nobreak\space}
\tx
\vglue 1cm
\centerline{\title    THE STELLAR INITIAL MASS FUNCTION
           \footnote{$^1$}{\tx Presented at the conference on ``Star
             Formation 1999'', Nagoya, Japan, June 21--25, 1999; to
                    be published, edited by T. Nakamoto}}
\space\space
\centerline{\author           Richard B. Larson}
\space
\centerline{              Yale Astronomy Department}
\centerline{            New Haven, CT 06520-8101, USA}
\centerline{                larson@astro.yale.edu}
\space\space\space
\centerline{                      ABSTRACT}
\space
{\narrower
   The current status of both the observational evidence and the theory
of the stellar initial mass function (IMF) is reviewed, with particular
attention to the two basic, apparently universal features shown by all
observations of nearby stellar systems: (1) a characteristic stellar
mass of the order of one solar mass, and (2) a power-law decline of the
IMF at large masses similar to the original Salpeter law.  Considerable
evidence and theoretical work supports the hypothesis that the
characteristic stellar mass derives from a characteristic scale of
fragmentation in star-forming clouds which is essentially the Jeans
scale as calculated from the typical temperature and pressure in
molecular clouds.  The power-law decline of the IMF at large masses
suggests that the most massive stars are built up by scale-free
accretion or accumulation processes, and the observed formation of
these stars in dense clusters and close multiple systems suggests that
interactions between dense prestellar clumps or protostars in forming
clusters will play a role.  A simple model postulating successive
mergers of subsystems in a forming cluster accompanied by the accretion
of a fraction of the residual gas by the most massive protostar during
each merger predicts an upper IMF of power-law form and reproduces the
Salpeter law with a plausible assumed accretion efficiency.
\space}

\section{1~~Introduction}

\noindent
   The stellar initial mass function (IMF), or distribution of masses
with which stars are formed, is the most fundamental output function
of the star formation process, and it controls nearly all aspects of
the evolution of stellar systems.  The importance of understanding
the origin of the IMF and its possible universality has therefore been
a stimulus for much research on star formation, both theoretical and
observational, and interest in this subject is of long standing, going
back at least to the pioneering study by Nakano (1966) of some of
the processes that might be responsible for determining the stellar
IMF\null.  In recent years there has been much progress in observational
studies relating to the IMF, and somewhat more modest progress in
reaching a theoretical understanding of its origin; here I review
briefly the current status of both the observational evidence and the
theoretical ideas concerning the origin IMF\null.  Other recent reviews
of the observations and the theory of the IMF have been given by Scalo
(1998), Clarke (1998), Larson (1998, 1999), Elmegreen (1999), and Meyer
et al.\ (2000).

\section{2~~Basic Observed Features of the Stellar IMF}

\noindent
   Numerous observational studies have been carried out to measure
or constrain the IMF in systems with as wide a range in properties as
possible in order to establish whether it is universal or whether it
varies with place or time, depending for example on parameters such as
metallicity.  The regions that have been studied with direct star counts
so far include the local field star population in our Galaxy and many
star clusters of all ages and metallicities in both our Galaxy and
the Magellanic Clouds.  As summarized below, this large body of direct
evidence does not yet demonstrate convincingly any variability of the
IMF, although the uncertainties are still large.  Some indirect evidence
based on the photometric properties of more distant and exotic systems
suggests that the IMF may vary in extreme circumstances, possibly being
more top-heavy in starbursts and high-redshift galaxies (Larson 1998),
but this indirect evidence is less secure and will not be discussed
further here.

   As reviewed by Miller \& Scalo (1979), Scalo (1986, 1998), Kroupa
(1998), and Meyer et al.\ (2000), the IMF derived for the field stars
in the solar neighborhood exhibits an approximate power-law decline with
mass above one solar mass that is consistent with, or somewhat steeper
than, the original Salpeter (1955) law; however, below one solar mass
the IMF of the field stars clearly flattens, showing a possible broad
peak at a few tenths of a solar mass in the number of stars per unit
logarithmic mass interval.  If the logarithmic slope $x$ of the IMF is
defined by $dN/d\log m \propto m^{-x}$, then the slope at large masses
is $x \sim 1.5$, while the slope at small masses is $x \sim 0$, the
range of values or uncertainty in $x$ being about $\pm 0.5$ in each
case.  The IMF inferred for the local field stars is subject to
significant uncertainty, especially in the range around one solar mass,
because it depends on the assumed evolutionary history of the local
Galactic disk and on assumed stellar lifetimes.  In contrast, the IMFs
of individual star clusters can be derived with fewer assumptions and
should be more reliable, since all of the stars in each cluster have the
same age and since, at least in the youngest clusters, all of the stars
ever formed are still present and can be directly counted as a function
of mass without the need for evolutionary corrections.  Much effort has
therefore gone into determining IMFs for clusters with a wide range of
properties in both our Galaxy and the Magellanic Clouds.  As reviewed
by von Hippel et al.\ (1996), Hunter et al.\ (1997), Massey (1998), and
Scalo (1998), the results of these studies are generally consistent with
the IMF inferred for the local field stars, and the values found for the
slope $x$ of the IMF above one solar mass generally scatter around the
Salpeter value $x = 1.35$ (see figure 5 of Scalo 1998).  In all cases in
which it has been possible to observe low-mass stars, the cluster IMFs
also show a flattening below one solar mass.  No clear evidence has been
found for any systematic dependence of the IMF on any property of the
systems studied, and this has led to the current widely held view that
the IMF is universal, at least in the local universe.

  Recent studies have provided more information about very faint stars
and brown dwarfs, and the IMF estimated for them remains approximately
flat or shows only a moderate decline into the brown dwarf regime,
consistent with an extrapolation of the IMF of lower main sequence stars
and showing no evidence for any abrupt truncation at low masses (Basri
\& Marcy 1997; Mart\'\i n et al.\ 1998; Bouvier et al.\ 1998; Reid
1998).  Another area of recent progress has been the determination of
IMFs for a number of newly formed star clusters that still contain
many pre-main-sequence stars; as reviewed by Meyer et al.\ (2000),
these results again show general consistency with the field star IMF,
including a similar flattening below one solar mass and a possible broad
peak at a few tenths of a solar mass.  Significant numbers of brown
dwarf candidates have been found in these young clusters, and although
the derivation of an IMF for them is complicated by the need to know
their ages accurately, the results again suggest an IMF that is flat or
moderately declining at the low end (Luhman \& Rieke 1999; Hillenbrand
\& Carpenter 1999).

  In summary, within the still rather large uncertainties, all of the
data that have been described are consistent with a universal IMF that
is nearly flat at low masses and that can be approximated by a declining
power law with a slope similar to the original Salpeter slope above one
solar mass.  The fact that the IMF cannot be approximated by a single
power law at all masses but flattens below one solar mass means that
there is a characteristic stellar mass of the order of one solar mass
such that most of the mass that condenses into stars goes into stars
with masses of this order.  In fact, a more robust statement about the
IMF than any claimed functional form is the fact that about 75\% of
the mass that forms stars goes into stars with masses between 0.1
and $10\,$M$_\odot$, while about 20\% goes into stars more massive
than $10\,$M$_\odot$ and only 5\% into stars less massive than
$0.1\,$M$_\odot$.  The existence of this characteristic stellar
mass is the most fundamental fact needing to be explained by any
theoretical understanding of star formation.  The second fundamental
fact to be explained is that a significant fraction of the mass goes
into massive stars in a power-law tail of the IMF extending to masses
much larger than the characteristic mass.  Possible theoretical
explanations of these two basic facts will be discussed in the
following sections.

\section{3~~The Origin of the Characteristic Stellar Mass}

\noindent
   For some years, there have been two contending viewpoints about the
origin of the characteristic stellar mass, one holding that it results
from a characteristic mass scale for the fragmentation of star-forming
clouds (e.g., Larson 1985, 1996), and the other holding that it results
from the generation of strong outflows at some stage of protostellar
accretion (e.g., Adams \& Fatuzzo 1996); both effects might in fact
play some role, as reviewed by Meyer et al.\ (2000).  The fragmentation
hypothesis for the origin of the characteristic mass has recently
received support from observations showing that the $\rho$~Ophiuchus
cloud contains many small, apparently pre-stellar clumps with masses
between 0.05 and $3\,$M$_\odot$ whose properties are consistent with
their having been formed by the gravitational fragmentation of the
cloud, and whose mass spectrum is very similar to the stellar IMF
discussed above, including the flattening below one solar mass (Motte,
Andr\'e, \& Neri 1998; see also Andr\'e 2000; Andr\'e, Ward-Thompson,
\& Barsony 2000).  In particular, the mass spectrum of the clumps in
the $\rho$~Oph cloud is quite similar to the mass spectrum of the young
stars observed in this cloud (Luhman \& Rieke 1999), suggesting that the
IMF of the stars derives directly from the mass spectrum of the clumps.
A clump mass spectrum consistent with the stellar IMF has also been
found in the Serpens cloud by Testi \& Sargent (1998), and additional
evidence for a possible mass scale of order one solar mass in the
structure of molecular clouds has been reviewed by Evans (1999) and
Williams, Blitz, \& McKee (2000).

   Although the original analysis of Jeans (1929) showing the existence
of critical length and mass scales for the fragmentation of a collapsing
cloud was not self-consistent, rigorous stability analyses that yield
dimensionally equivalent results can be made for various equilibrium
configurations, including sheets, disks, and filaments (Spitzer 1978;
Larson 1985).  In all cases, there is a predicted characteristic mass
scale for fragmentation that is a few times $c^4\!/G^2\mu$, where
$c$ is the isothermal sound speed and $\mu$ is the surface density
of the assumed equilibrium configuration.  For a typical molecular
cloud temperature of $10\,$K and a typical surface density of
100~M$_\odot\,$pc$^{-2}$, this mass scale is about one solar mass,
similar to the observed typical stellar mass (Larson 1985).  Alternatively, if collapsing pre-stellar clumps form not by the
fragmentation of equilibrium configurations but as condensations in
a medium with some characteristic ambient pressure $P$, the minimum
mass that can collapse gravitationally is that of a marginally
stable `Bonnor-Ebert' sphere with a boundary pressure $P$, or
$1.18\,c^4\!/G^{3/2}P^{1/2}$ (Spitzer 1968).  Since any self-gravitating
configuration has an internal pressure $P \sim \pi G\mu^2\!/2$, this
result is dimensionally equivalent to the fragmentation scale given
above, and it can be regarded as a different expression for the same
basic physical quantity, which can still conveniently be called the
`Jeans mass'.

   It is not yet clear to what extent star-forming molecular clouds or
their denser subregions can be regarded as equilibrium configurations,
and it may instead be that much of the structure in these clouds
consists of transient density fluctuations generated by supersonic
turbulence (Larson 1981).  Some of the filamentary structure in
molecular clouds may be created by violent dynamical phenomena in an
active star-forming environment (Bally et al.\ 1991), and simulations
of turbulence in the interstellar medium often show the appearance
of transient filamentary features that form where supersonic
turbulent flows converge (V\'azquez-Semadeni et al.\ 1995, 2000;
Ballesteros-Paredes et al.\ 1999).  In the presence of gravity, some of
the densest clumps produced in this way may become self-gravitating and
begin to collapse; the initial state for their collapse might then be
roughly approximated by a marginally stable Bonnor-Ebert sphere whose
boundary pressure is determined by the ram pressure of the turbulent
flow.  If there is a rough balance between turbulent pressure
and gravity in molecular clouds, the turbulent pressure will be
approximately equal to the gravitational pressure $P \sim \pi
G\mu^2\!/2$, yielding a pressure $P \sim 3 \times 10^5$ cm$^{-3}\,$K
for a typical surface density $\mu \sim 100$~M$_\odot\,$pc$^{-2}$.
Alternatively, a typical pressure may be estimated by noting that the
correlations among linewidth, size, and density that hold among many
molecular clouds (Larson 1981; Myers \& Goodman 1988) imply that these
clouds all have similar turbulent ram pressures $\rho v^2$, for which a
typical value is again approximately $3 \times 10^5$ cm$^{-3}\,$K\null.
For a marginally stable Bonnor-Ebert sphere with a temperature of 10~K
bounded by this pressure, the predicted mass and radius are about
$0.7\,$M$_\odot$ and $0.03\,$pc respectively (Larson 1991, 1996, 1999).
Although factors of 2 may not be very meaningful, these quantities are
similar in magnitude to the typical masses and sizes of the pre-stellar
clumps observed in molecular clouds (e.g., Motte et al.\ 1998) and to
the characteristic stellar mass noted above.  Thus there may indeed be
an intrinsic mass scale in the star formation process, and this mass
scale may be essentially the Jeans mass as defined above.

  There are also some hints that there may be a corresponding size
scale for star-forming clumps.  Analyses of the spatial distributions
of the newly formed T~Tauri stars in several regions show the existence
of two regimes in a plot of average companion surface density versus
separation, namely a binary regime with a steep slope at small
separations and a clustering regime with a shallower slope at large
separations, with a clear break between them at a separation of
about $0.04\,$pc that may represent the size of a typical collapsing
pre-stellar clump (Larson 1995; Simon 1997).  Although this
interpretation of the observations is not unique and the scale of
the break may also depend on superposition effects and on the dynamical
evolution of the system (Nakajima et al.\ 1998; Bate, Clarke, \&
McCaughrean 1998), the interpretation of the break in terms of a typical
clump size may still be valid in low-density regions like Taurus where
these effects may not be as important as in denser regions.  A similar
size scale has been found by Ohashi et al.\ (1997) in a study of the
rotational properties of collapsing pre-stellar clumps, which shows
that their specific angular momentum is apparently conserved on scales
smaller than about $0.03\,$pc; this may represent the characteristic
size of a region that collapses rapidly to form a star or binary system
(see also Ohashi 2000; Myers, Evans, \& Ohashi 2000).  Finally, an
analysis of the internal kinematics of star-forming cloud cores by
Goodman et al.\ (1998) shows a transition from a turbulent regime on
large scales, where the linewidth increases systematically with region
size, to a regime of `velocity coherence' on scales smaller than about
$0.1\,$pc, where the linewidth becomes nearly independent of region
size.  These authors suggest that this change in kinematic behavior is
related to the break between the clustering and binary regimes for the
T~Tauri stars noted above, and they suggest that it has the same basic
cause, namely a transition from chaotic dynamics on large scales to more
ordered behavior on small scales.  Such a transition might be expected
because molecular clouds are dominated by turbulent and magnetic
pressures on large scales and by thermal pressure on small scales
(Larson 1981; Myers 1983), and the transition between the two regimes
is in fact what defines the Jeans scale when the latter is calculated by
assuming pressure balance between a thermally supported isothermal clump
and a turbulent ambient medium.  All of the evidence described here is
thus consistent with the existence of a scale in the star formation
process which is essentially the Jeans scale as derived above.

   The characteristic stellar mass may thus depend, via the Jeans mass,
on the typical temperature and pressure in star-forming clouds, being
proportional to $T^2\!/P^{1/2}$.  The temperatures of molecular clouds
are controlled by radiative processes, but their pressures are probably
of dynamical origin and result from the cloud formation process, since
their internal pressures are much higher than the general pressure of
the interstellar medium (Larson 1996).  Molecular clouds are probably
created by the collisional agglomeration of smaller, mostly atomic
clouds in regions where large-scale converging flows assemble the atomic
clouds into large complexes.  The resulting cloud collisions produce
a ram pressure $\rho v^2$ which may determine the typical internal
pressure of the molecular clouds formed.  If the typical density of
the colliding clouds is 20 atoms per cm$^3$ and if they collide with
a velocity of 10 km$\,$s$^{-1}$, the ram pressure produced is $\sim 3
\times 10^5$ cm$^{-3}\,$K, similar to the inferred internal pressures
of molecular clouds.  Thus the typical pressures in molecular clouds
can be understood in terms of the structure and dynamics of the atomic
component of the interstellar medium.  It may further be possible to
understand the properties of the atomic clouds in terms of the classical
two-phase model of the ISM, which postulates a balance in thermal
pressure between a cool cloud component and a warm intercloud component
and predicts cloud densities of a few tens of atoms per cm$^3$ (Field,
Goldsmith, \& Habing 1969; Wolfire et al.\ 1995).  Thus it may be
possible to understand the characteristic temperatures and pressures
of molecular clouds, and hence the characteristic stellar mass, in
terms of relatively well-studied thermal and dynamical properties of
the interstellar medium (Larson 1996).

   If the mass scale for star formation depends on the temperature and
pressure in star-forming clouds as predicted above, one might expect
to see some variability of the IMF between regions with different
properties; for example, clouds with higher temperatures might be
expected to form stars with a higher characteristic mass (Larson 1985).
There is possible evidence for such an effect in extreme cases such
as starburst systems and high-redshift galaxies (Larson 1998), but no
clear dependence of the IMF on the temperature or other properties of
star-forming clouds has been found in local star-forming regions.  In
fact, clouds with higher temperatures generally also have much higher
pressures, so there is a partial cancellation of these effects when the
Jeans mass is calculated, and it is not clear that one effect or the
other dominates.  Elmegreen (1999) has argued that such an approximate
cancellation of effects is to be expected for physical reasons since the
cloud temperature depends on radiative heating rates while the overall
pressure of the ISM depends on the local column density of matter in a
galaxy, both of which increase with the stellar surface density in such
a way that $T^2\!/P^{1/2}$ is approximately constant.

\section{4~~The Formation of Massive Stars and the Origin of the
            Power-Law Upper IMF}

\noindent
   The second basic fact about star formation needing to be explained
is that the IMF has a power-law tail extending to masses much larger
than the characteristic mass, such that about 20\% of the total mass
goes into stars more massive than $10\,$M$_\odot$.  Most of the feedback
effects of star formation on the evolution of galaxies depend on
energy input from these massive stars, so it is clearly of great
importance to understand the origin and possible universality of the
upper IMF\null.  At present the formation of massive stars is relatively
poorly understood, both observationally and theoretically, so most of
what can be said about the origin of the upper IMF remains speculative.
Recent observational and theoretical progress in understanding the
formation of massive stars has been reviewed by Evans (1999), Garay
\& Lizano (1999), and Stahler, Palla, \& Ho (2000).

   A theoretical constraint on the formation processes of massive stars
is provided by the fact that, for stellar masses larger than about
$10\,$M$_\odot$, radiation pressure begins to exceed gravity in the
infalling envelope around an accreting protostar (Wolfire \& Cassinelli
1987); this means that standard radial infall models probably cannot
account for the formation of stars much more massive than about
$10\,$M$_\odot$, although such models may still suffice for stars of
up to about this mass (Stahler et al.\ 2000).  Therefore, non-spherical
or non-uniform accretion processes are probably required to continue
building up the most massive stars.  One possibility is that the
infalling gas settles into a disk which can then be accreted without
hindrance from radiation pressure (Nakano 1989; Jijina \& Adams 1996).
Evidence that disks may play a role in the formation of massive stars
has been reviewed by Garay \& Lizano (1999), but the role of disks
for massive stars is not as clear as in the case of low mass stars.
Another possibility is that the formation of massive stars involves
the accretion of very dense clumps, or of dense circumstellar matter
accreted as a result of interactions among protostars in a forming
cluster of stars (Larson 1982, 1990).

   Relevant observational evidence is provided by the fact that newly
formed massive stars are always found to be surrounded by clusters
of less massive stars, the more massive stars tending to have larger
associated clusters (Hillenbrand 1995; Testi, Palla, \& Natta 1999;
Garay \& Lizano 1999).  This means that the conditions that favor the
formation of massive stars also favor the formation of many less massive
stars in the same vicinity.  The most massive stars in young clusters
tend to be centrally located in these clusters, as is exemplified by
the Trapezium system (Larson 1982; Zinnecker, McCaughrean, \& Wilking
1993; Hillenbrand \& Hartmann 1998), and this can be understood only if
these stars were in fact formed near the cluster center (Bonnell \&
Davies 1998).  Massive stars also have a high frequency of massive
companions, and even the runaway O~stars must have been formed in close
proximity to other massive stars in very dense stellar systems (Stahler
et al.\ 2000).  All of this evidence indicates that massive stars form
only in regions of exceptionally high density along with many less
massive stars, and that they typically form in very close proximity
to other massive stars.  Massive star-forming cloud cores also show
evidence for more internal substructure than the less massive cores that
have been more widely studied (Evans 1999).  Interactions among the many
dense pre-stellar clumps and accreting protostars that must exist in
such an environment will therefore almost certainly play a role in the
accretional growth of the most massive stars, perhaps accounting for
the accretion by them of matter that is sufficiently dense to overcome
the effects of radiation pressure.  Large amounts of matter must be
accumulated very rapidly to form a massive star, so the process must
be a rather violent one.  An extreme case of such a violent formation
process, which must sometimes happen, would be the merging of two
already-formed less massive stars (Bonnell, Bate, \& Zinnecker 1998;
Stahler et al.\ 2000).

   If no new mass scale larger than the Jeans mass enters the problem,
it is possible that the accumulation processes involved in the formation
of the massive stars might proceed in an approximately scale-free
fashion to build up a power-law upper IMF\null.  Several types of
approximately scale-free accumulation models have been considered in
efforts to explain how a power-law upper IMF might be produced.  The
first to be developed in some detail was that of Nakano (1966), who
suggested that clumps formed by the fragmentation of a collapsing cloud
would collide randomly and sometimes coalesce to create a spectrum of
clump masses extending to values much larger than the initial fragment
mass; he showed that this process could yield an approximate power-law
mass spectrum similar to the observed IMFs of some star clusters.
Such models were elaborated further by Arny \& Weissman (1973), Silk
\& Takahashi (1979), Pumphrey \& Scalo (1983), and Nakano, Hasegawa, \&
Norman (1995).  A second possibility is that protostars might continue
to accrete ambient matter gravitationally at a rate that increases with
their mass, as is true for Bondi-Hoyle accretion whose rate increases
with the square of the mass; this process can build up a power-law tail
on the IMF with a slope $x = 1$ (Zinnecker 1982).  A third possibility
is that if stars form in a hierarchy of groups and subgroups, and if
accumulation processes tend to build more massive stars in the more
massive subgroups of such a hierarchy, then a power-law upper IMF
can be produced (Larson 1991, 1992).  Most stars do indeed form in
clusters, and in at least some cases there is evidence for hierarchical
subclustering (Zinnecker et al.\ 1993; Gomez et al.\ 1993; Larson 1995;
Elmegreen et al.\ 2000).  Since the larger subgroups in such a hierarchy
contain more `raw materials' from which to build massive stars, they
will almost certainly produce stars with a mass spectrum extending to a
larger maximum mass.  If the mass $M_{\rm max}$ of the most massive star
formed in any subgroup increases with a power $n < 1$ of the mass of the
subgroup, i.e.\ if $M_{\rm max} \propto M_{\rm group}^n$, and if all
stars form in a self-similar hierarchy of such groups, then a power-law
IMF is generated whose slope is $x = 1/n$ (Larson 1992). For example,
the IMF slope $x = 1.4 \pm 0.4$ suggested by the evidence discussed in
Section~2 could be reproduced if $n$ were $0.7 \pm 0.2$.

   One hypothesis involving hierarchical structure that has been
developed further is that star-forming clouds have fractal structures,
and that the universal power-law upper IMF results from a universal
fractal cloud structure produced by turbulence (Larson 1992, 1995;
Elmegreen 1997, 1999).  In the model of Larson (1992), stars are assumed
to form by gas accumulation along filaments in a fractal filamentary
network, and the resulting IMF slope $x$ is equal to the fractal
dimension $D$ of the network.  Elmegreen (1997) has proposed a more
generic model in which stars form by random selection from different
levels of any fractal hierarchy.  However, while there is evidence that
molecular clouds have fractal {\it boundary shapes}, it is less clear
that they have fractal {\it mass distributions}, and most of their mass
cannot plausibly be fractally distributed but must have a smoother
spatial distribution.  In any case, even if a fractal picture were
correct, the accumulation processes required to form stars in such a
model would first form small stars from small cloud substructures before
matter could be accumulated from larger regions to form more massive
stars; the cloud regions that form massive stars would then contain
substructure that has already begun to form less massive stars. Such a
picture would predict the formation of massive stars only in clusters,
as is indeed observed, but the interactions among star-forming clumps
and protostars that would necessarily occur during the accumulation of
matter to form the more massive stars were not taken into account in the
above fractal models.  Such interactions would almost certainly play a
role in determining the final stellar mass spectrum.

   It may in fact be that all of the ideas mentioned above have some
merit, and that a more realistic model will involve elements of all
of them, namely clump collisions, continuing gas accretion, and
hierarchical clustering.  In its original form, the clump coagulation
model of Nakano (1966) did not take into account the fact that clumps
formed by the fragmentation of a contracting cloud will often begin to
collapse into stars before colliding and interacting with each other.
Many of the colliding clumps will then contain accreting protostars,
and the effects of their interactions on the protostellar accretion
process and on the structure of the forming system of stars will play
an important role in its further development.  Since these interactions
will generally be dissipative, the star-forming clumps will tend to
become bound into progressively larger and denser aggregates (Larson
1990).  In this way, star clusters may be built up hierarchically by
the merging of smaller subsystems, perhaps basically as in the clump
coagulation model of Nakano (1966) but with the clumps replaced here
by groups of forming stars.  For a brief time, a newly formed cluster
of stars may continue to show hierarchical subclustering, but this
substructure will soon be erased by dynamical relaxation processes.
As smaller systems of forming stars continue to merge into larger ones,
the protostars in the most favored central locations may continue to
gain mass from larger and larger accretion zones, building up an
extended spectrum of stellar masses.

   Numerical simulations illustrate the likely importance of
interactions for the continuing accretional growth of the more massive stars in such a scenario.  Interactions between newly formed stars with
residual disks can strongly perturb their surrounding disks, causing
part of the disk matter to be ejected and part to be accreted by the
central star (Heller 1991, 1995); in general, the more massive system
tends to gain mass from the less massive one in such interactions.  In
the simulations of cloud fragmentation and accretion by Larson (1978),
the most massive objects gained much of their final mass during episodes
of rapid accretion associated with close encounters or mergers between
dense clumps.  Simulations of accretion processes in forming clusters
of stars (Bonnell et al.\ 1997, 1998; Clarke, Bonnell, \& Hillenbrand
2000) show the development of a broad spectrum of masses, the more
massive objects tending to form near the cluster center where the
accretion and interaction rates are highest; the most massive stars may
even gain much of their final mass by mergers between already-formed
stars (Bonnell et al.\ 1998; Stahler et al.\ 2000).

   The simple Bondi-Hoyle accretion model of Zinnecker (1982) assumes
a protostellar accretion rate that increases with mass in qualitatively
the expected way, and it predicts a rapidly increasing spread in
protostellar masses and the growth of a power-law tail on the IMF that
is qualitatively similar to what is observed.  However, it also has
the unrealistic feature that it predicts the unlimited runaway growth
in mass of the most massive protostar because it is assumed to accrete
matter from a region of unlimited size.  More realistically, each
protostar in a forming cluster will have an accretion zone of finite
size associated with the subsystem in which it forms (Larson 1978), and
the total amount of gas available to form massive stars will be limited
by the size of the cluster.  Since the gas supply is depleted as
accreting protostars continue to gain mass from it, a decreasing amount
of mass is available to build stars of higher and higher mass, resulting
in an IMF with $x > 1$ in which there is less and less mass in stars of
increasing mass, as is observed.  The amount of mass accreted by each
protostar may then be determined by the amount of gas in the subsystem
in which it forms, and by the effects of continuing interactions and
mergers among the subsystems in a forming cluster; each such interaction
or merger is likely to cause additional gas to be accreted by the most
massive protostar present.

   One can easily construct simple interaction and accretion schemes
based on these ideas that generate a power-law IMF\null.  The only
essential requirement is that the accretion processes involved are
basically scale-free, that is, they do not depend on any new mass scale
larger than the Jeans mass.  This would be the case if, for example,
each interaction or merger between two subsystems causes a constant
fraction of the remaining gas to be accreted by the most massive
protostar present.  If we assume, in the simplest formulation of such
a model, that the mass of the most massive protostar increases by a
constant factor $f$ when the mass of the system to which it belongs
increases by another constant factor $g$ because of a merger with
another system (for example, $g = 2$ for equal-mass mergers), then the
mass of the most massive star formed in a cluster built up by a sequence
of such mergers increases as a power $n$ of the cluster mass, where
$n = \log f/\log g$. If all stars more massive than the Jeans mass are
formed in a self-similar hierarchy of such merging subsystems, then the
assumptions of the hierarchical clustering model of Larson (1991, 1992)
are satisfied and a power-law upper IMF is produced that has a slope
$x = 1/n$.  The Salpeter slope is recovered if, for example, $g = 2$
and $f = 5/3$; then $n = \log (5/3)/\log 2 = 0.74$ and $x = 1/n = 1.36$.
If the most massive protostar grows by accreting residual gas, then it
can be shown that in this simple example, 1/6 of the remaining gas in
the two subsystems is accreted during each merger.  Conversely, if it is
assumed that 1/6 of the remaining gas is accreted by the most massive
protostar during each merger, then a Salpeter IMF is produced.  If the
fraction of the gas accreted in each merger varies between 1/10 and 1/4,
then the predicted value of $x$ varies between 1.18 and 1.71.  These
results are not very sensitive to the assumption of equal-mass mergers;
for example, if the mass ratio of the interacting subsystems is not 1
but 3, a typical value for clump coalescence models, and if again 1/6
of the remaining gas is accreted during each merger, the resulting
IMF slope is $x = 1.44$.  These assumptions do not seem obviously
implausible in the light of the observational evidence and the
theoretical results noted above, and they all result in IMF slopes
that are consistent with the observations, within the uncertainties.

   While it would be easy to construct more elaborate and perhaps more
realistic accretion models that also yield power-law IMFs, what is
really needed to advance our understanding of the origin of the upper
IMF is better physical input regarding the processes involved in
the accretional growth of massive stars, and estimates of the efficiency
of these processes, for example the fraction of residual gas accreted
during each interaction or merger between subsystems.  The processes
involved can now be studied in some detail with numerical simulations,
which as noted above have already begun to simulate some of the
processes likely to be important.  At present these simulations do
not provide sufficient quantitative information to test in any detail
the kind of model that has been proposed.  However, if more detailed
simulations support the kind of interaction/accretion picture suggested
above, and if the accretion processes involved are indeed approximately
scale-free and characterized by similar efficiencies, important progress
will have been made toward understanding the formation of massive stars
and the origin of the upper IMF\null.  Ultimately such simulations will
have to reproduce not only the IMF of the massive stars but also the
clustering and binary properties of these stars as well, and this test
will place strong constraints on the models.  Whatever processes may be
involved, the formation of massive stars cannot be understood without
explaining the striking facts that they form only in dense clusters,
and typically in very close proximity to other massive stars.  It seems
almost unavoidable that complex and perhaps violent dynamical
interactions will play an important role.

\section{References}

{\leftskip=5mm \parindent=-5mm

Adams, F. C., \& Fatuzzo, M. 1996, ApJ, 464, 256

Andr\'e, P. 2000, this conference

Andr\'e, P., Ward-Thompson, D., \& Barsony, M. 2000, in
   Protostars and Planets IV, eds.\ V. Mannings, A.~P. Boss, \& S.~S.
   Russell (University of Arizona Press, Tucson), in press

Arny, T., \& Weissman, P. 1973, AJ, 78, 309

Ballesteros-Paredes, J., V\'azquez-Semadeni, E., \& Scalo,
   J. 1999, ApJ, 515, 286

Bally, J., Langer, W. D., Wilson, R. W., Stark, A.~A., \&
   Pound, M.~W.. 1991, in Fragmentation of Molecular Clouds and Star
   Formation, IAU Symposium No.\ 147, eds.\ E. Falgarone, F. Boulanger,
   \& G. Duvert (Kluwer, Dordrecht), 11

Basri, G., \& Marcy, G. W. 1997, in Star Formation Near
   and Far, eds.\ S.~S. Holt \& L.~G. Mundy (AIP Conference Proceedings
   393, Woodbury, NY), 228 

Bate, M. R., Clarke, C. J., \& McCaughrean, M.~J. 1998,
   MNRAS, 297, 1163

Bonnell, I. A., \& Davies, M. B. 1998, MNRAS, 295, 691

Bonnell, I. A., Bate, M. R., Clarke, C.~J., \& Pringle,
   J.~E. 1997, MNRAS, 285, 201

Bonnell, I. A., Bate, M. R., \& Zinnecker, H., 1998, MNRAS,
   298, 93

Bouvier, J., Stauffer, J.R., Mart\'\i n, E.L., Barrado y
   Navascu\'es, D., Wallace, B., \& B\'ejar, V.~J.~S. 1998, A\&A, 336,
   490

Clarke, C. 1998, in The Stellar Initial Mass Function,
   eds.\ G. Gilmore and D. Howell (ASP Conference Series, Vol. 142,
   San Francisco), 189

Clarke, C. J., Bonnell, I. A., \& Hillenbrand, L.~A. 2000,
   in Protostars and Planets IV, eds.\ V. Mannings, A.~P. Boss, \&
   S.~S. Russell (University of Arizona Press, Tucson), in press

Elmegreen, B. G. 1997, ApJ, 486, 944

Elmegreen, B. G. 1999, in The Evolution of Galaxies on
   Cosmological Timescales, eds.\ J.~E. Beckman \& T.~J. Mahoney (ASP
   Conference Series, San Francisco), in press

Elmegreen, B. G., Efremov, Y., Pudritz, R.~E., \& Zinnecker,
   H. 2000, in Protostars and Planets IV, eds.\ V. Mannings, A.~P. Boss,
   \& S.~S. Russell (University of Arizona Press, Tucson), in press

Evans, N. J. 1999, ARA\&A, 37, in press

Field, G. B., Goldsmith, D. W., \& Habing, H.~J. 1969, ApJ,
   155, L149

Garay, G., \& Lizano, S. 1999, PASP, in press

Gomez, M., Hartmann, L., Kenyon, S. J., \& Hewett, R. 1993,
   AJ, 105, 1927

Goodman, A. A., Barranco, J. A., Wilner, D.~J., \& Heyer,
   M.~H. 1998, ApJ, 504, 223

Heller, C. H. 1991, PhD thesis, Yale University

Heller, C. H. 1995, ApJ, 455, 252

Hillenbrand, L. A. 1995, PhD thesis, Univ.\ of Massachusetts 

Hillenbrand, L., \& Carpenter, J. 1999, BAAS, 31, 906

Hillenbrand, L. A., \& Hartmann, L. W. 1998, ApJ, 492, 540

Hunter, D. A., Light, R. M., Holtzman, J. A., Lynds, R.,
   O'Neil, E.~J., \& Grillmair, C.~J. 1997, ApJ, 478, 124 

Jeans, J. H. 1929, Astronomy and Cosmogony (Cambridge
   University Press, Cambridge; reprinted by Dover, New York, 1961)

Jijina, J., \& Adams, F. C. 1996, ApJ, 462, 874

Kroupa, P. 1998, in Brown Dwarfs and Extrasolar Planets,
   eds.\ R. Rebolo, E.~L. Mart\'\i n, \& M.~R. Zapatero-Osorio (ASP
   Conference Series, Vol.\ 134, San Francisco), 483

Larson, R. B. 1978, MNRAS, 184, 69

Larson, R. B. 1981, MNRAS, 194, 809

Larson, R. B. 1982, MNRAS, 200, 159

Larson, R. B. 1985, MNRAS, 214, 379

Larson, R. B. 1990, in Physical Processes in Fragmentation
   and Star Formation, eds.\ R. Capuzzo-Dolcetta, C. Chiosi, \& A. Di
   Fazio (Kluwer, Dordrecht), 389

Larson, R. B. 1991, in Fragmentation of Molecular Clouds
   and Star Formation, IAU Symposium No.\ 147, eds.\ E. Falgarone, F.
   Boulanger, \& G. Duvert (Kluwer, Dordrecht), 261

Larson, R. B. 1992, MNRAS, 256, 641

Larson, R. B. 1995, MNRAS, 272, 213

Larson, R. B. 1996, in The Interplay Between Massive Star
   Formation, the ISM and Galaxy Evolution, eds.\ D. Kunth, B.
   Guiderdoni, M. Heydari-Malayeri, \& T.X. Thuan (Editions
   Fronti\`eres, Gif sur Yvette), 3 

Larson, R. B. 1998, MNRAS, 301, 569

Larson, R. B. 1999, in The Orion Complex Revisited,
   eds.\ M.~J. McCaughrean \& A. Burkert (ASP Conference Series, San
   Francisco), in press

Luhman, K. L., \& Rieke, G. H. 1999, ApJ, in press

Mart\'\i n, E. L., Zapatero-Osorio, M. R., \& Rebolo, R.
   1998, in Brown Dwarfs and Extrasolar Planets, eds.\ R. Rebolo, E.~L.
   Mart\'\i n, \& M.~R. Zapatero-Osorio (ASP Conference Series, Vol.\
   134, San Francisco), 507

Massey, P. 1998, in The Stellar Initial Mass Function,
   eds.\ G. Gilmore \& D. Howell (ASP Conference Series, Vol.\ 142,
   San Francisco), 17

Meyer, M. R., Adams, F. C., Hillenbrand, L.~A., Carpenter,
   J.~M., \& Larson, R.~B. 2000, in Protostars and Planets IV, eds.\
   V. Mannings, A.~P. Boss, \& S.~S. Russell (University of Arizona
   Press, Tucson), in press 

Miller, G. E., \& Scalo, J. M. 1979, ApJS, 41, 513

Motte, F., Andr\'e, P., \& Neri, R. 1998, A\&A, 336, 150

Myers, P. C. 1983, ApJ, 270, 105

Myers, P. C., \& Goodman, A. A. 1988, ApJ, 329, 392

Myers, P. C., Evans, N. J., \& Ohashi, N. 2000, in
   Protostars and Planets IV, eds.\ V. Mannings, A.~P. Boss, \& S.~S.
   Russell (University of Arizona Press, Tucson), in press

Nakajima, Y., Tachihara, K., Hanawa, T., Nakano, M. 1998,
   ApJ, 497, 721  

Nakano, T. 1966, Prog.\ Theor.\ Phys., 36, 515

Nakano, T. 1989, ApJ, 345, 464

Nakano, T., Hasegawa, T., \& Norman, C. 1995, ApJ, 450, 183

Ohashi, N. 2000, this conference

Ohashi, N., Hayashi, M., Ho, P.~T.~P., Momose, M., Tamura,
   M., Hirano, N.  \& Sargent, A.~I. 1997, ApJ, 488, 317

Pumphrey, W. A., \& Scalo, J. M. 1983, ApJ, 269, 531

Reid, I. N. 1998, in The Stellar Initial Mass Function,
   eds.\ G. Gilmore \& D. Howell (ASP Conference Series, Vol.\ 142, San
   Francisco), 121

Salpeter, E. E. 1955, ApJ, 121, 161

Scalo, J. 1986, Fundam.\ Cosmic Phys., 11, 1

Scalo, J. 1998, in The Stellar Initial Mass Function, eds.\
   G. Gilmore \& D. Howell (ASP Conference Series, Vol.\ 142, San
   Francisco), 201

Silk, J., \& Takahashi, T. 1979, ApJ, 229, 242

Simon, M. 1997, ApJ, 482, L81

Spitzer, L. 1968, in Nebulae and Interstellar Matter (Stars
   and Stellar Systems, Vol.\ 7), eds.\ B.~M. Middlehurst \& L.~H. Aller
   (University of Chicago Press, Chicago), 1

Spitzer, L. 1978, Physical Processes in the Interstellar
   Medium (Wiley-Interscience, New York)

Stahler, S. W., Palla, F., \& Ho, P.~T.~P. 2000, in
   Protostars and Planets IV, eds.\ V. Mannings, A.~P. Boss, \& S.~S.
   Russell (University of Arizona Press, Tucson), in press

Testi, L., \& Sargent, A.~I. 1998, ApJ, 508, L91

Testi, L., Palla. F., \& Natta, A. 1999, A\&A, 342, 515

V\'azquez-Semadeni, E., Passot, T., \& Pouquet, A., 1995,
   ApJ, 441, 702

V\'azquez-Semadeni, E., Ostriker, E.~C., Passot, T., Gammie,
   C.~F., \& Stone, J.~M. 2000, in Protostars and Planets IV, eds.\ V.
   Mannings, A.~P. Boss, \& S.~S.  Russell (University of Arizona Press,
   Tucson), in press

von Hippel, T., Gilmore, G., Tanvir, N., Robinson, D., \&
   Jones, D.~H.~P. 1996, AJ, 112, 192

Williams, J. P., Blitz, L., \& McKee, C.~F. 2000, in
   Protostars and Planets IV, eds.\ V. Mannings, A.~P. Boss, \& S.~S.
   Russell (University of Arizona Press, Tucson), in press

Wolfire, M. G., \& Cassinelli, J. P. 1987, ApJ, 319, 850

Wolfire, M. G., Hollenbach, D., McKee, C.~F., Tielens, A.~G.
   G.~M., \& Bakes, E.~L.~O. 1995, ApJ, 443, 152

Zinnecker, H. 1982, in Symposium on the Orion Nebula to
   Honor Henry Draper, eds.\ A.~E. Glassgold, P.~J. Huggins, \& E.~L.
   Schucking (Ann.\ New York Academy of Sciences, Vol. 395, New York),
   226

Zinnecker, H., McCaughrean, M. J., \& Wilking, B.~A. 1993,
   in Protostars and Planets III, eds.\ E.~H. Levy \& J.~I. Lunine
   (University of Arizona Press, Tucson), 429

}
\bye